\documentclass[submission,copyright,creativecommons]{eptcs}
\providecommand{\event}{HCVS 2016} 
\usepackage{breakurl}             

\usepackage{flushend}
\usepackage{amssymb,amsmath}

\usepackage{pifont}

\usepackage{xcolor}
\usepackage{paralist}
\usepackage{url}
\usepackage{pgfplots}
\usepackage{xspace}
\usepackage[normalem]{ulem}

\usepackage{wrapfig}
\usepackage{subcaption}
\usepackage{tikz}
\usetikzlibrary{arrows,decorations,shapes,shadows,calc}

\newcommand{\anony}[1]{}

\newcommand{\pponly}[1]{} 

\usepackage[linesnumbered,ruled,vlined]{algorithm2e}
\SetKw{KwLet}{let}
\SetKw{KwGlobal}{global}
\SetKw{KwAbort}{abort}
\SetKwFor{Forall}{for all}{do}{end}
\SetKwFor{Func}{function}{}{}
\SetKw{KwChoose}{choose}

\usepackage{listings}
\renewcommand{\vec}[1]{{\boldsymbol{#1}}}

\newcommand{\x}{{x}} 

\newcommand{\xargin}{{\x^{p\_in}}}
\newcommand{\xargout}{{\x^{p\_out}}}
\newcommand{\vx}{{\vec{\x}}}
\newcommand{\vxp}{{\vec{\x}'}}

\newcommand{\vxin}{{\vx^{in}}}

\newcommand{\vxout}{{\vx^{out}}}
\newcommand{\vxargin}{{\vx^{p\_in}}}
\newcommand{\vxargout}{{\vx^{p\_out}}}

\newcommand{\rrank}{RR}

\newcommand{\trans}{\mathit{Trans}}
\newcommand{\inv}{\mathit{Inv}}

\newcommand{\init}{\mathit{Init}}
\newcommand{\fout}{\mathit{Out}}

\newcommand{\precond}{\mathit{Precond}}

\newcommand{\summary}{\mathit{Sum}}
\newcommand{\ssummary}{\mathit{Summary}}

\newcommand{\callctx}{\mathit{CallCtx}}

\renewcommand{\function}{{procedure}}
\newcommand{\functions}{{procedures}}

\renewcommand{\paragraph}[1]{\smallskip\noindent\textbf{#1}\quad}

\usepackage{tikz}
\usetikzlibrary{positioning,shapes.geometric}
\usetikzlibrary{calc}
\usetikzlibrary{arrows,shapes,shadows,decorations}
\usetikzlibrary{decorations.pathreplacing}
\tikzset{
  ashadow/.style={opacity=.25, shadow xshift=0.07, shadow yshift=-0.07},
}
\usetikzlibrary{shapes.multipart}
\usetikzlibrary{patterns}
\usepackage{pgfplots}

\usepackage{pstricks}
\newrgbcolor{lightblue}{0.6 1.0 0.8}
\newrgbcolor{lightyellow}{1.0 1.0 0.4}
\newrgbcolor{darkgreen}{0 0.5 0}
\newrgbcolor{darkred}{0.5 0 0}

\title{Challenges in Decomposing\\ Encodings of Verification Problems}
\author{Peter Schrammel
\institute{University of Oxford, UK}
\email{peter.schrammel@cs.ox.ac.uk}
}
\def\titlerunning{Challenges in Decomposing Encodings of Verification Problems}
\def\authorrunning{Peter Schrammel}
\begin{document}
\maketitle

\begin{abstract}
Modern program verifiers use logic-based encodings of the verification
problem that are discharged by a back end reasoning engine.  However,
instances of such encodings for large programs can quickly overwhelm
these back end solvers.  Hence, we need techniques to make the solving
process scale to large systems, such as partitioning
(divide-and-conquer) and abstraction.
In recent work, we showed how decomposing the formula encoding of a
termination analysis can significantly increase efficiency.  The
analysis generates a sequence of logical formulas with existentially
quantified predicates that are solved by a synthesis-based program
analysis engine.  However, decomposition introduces abstractions in
addition to those required for finding the unknown predicates in the
formula, and can hence deteriorate precision. We discuss the
challenges associated with such decompositions and their
interdependencies with the solving process.
\end{abstract}

\section{Introduction}
\vspace*{-0.9ex}

Logic-based encodings of the verification problem 
are more and more widespread in
software verification~\cite{BGMR15}.
However, the generated formulae are often too large to be 
directly handled by the back end solver. Classical divide-and-conquer
techniques suggest themselves to cope with such large problems.
Work on interprocedural verification, e.g.\ \cite{KGC14,CDK+15},
follows the syntactical, procedural structure of the program to perform
a decomposition of the formula.
This does not seem ideal, but has been shown to significantly increase
efficiency in comparison with monolithic solving.

In recent work, we used a synthesis engine \cite{SK16,BJKS15} to solve
for multiple predicates at once even when they are mutually dependent.
Since this scales badly to large formulae, we have to
decompose the formula in order to reduce the load on the synthesis
engine.
However, the decomposition may introduce additional abstractions,
in particular when mutually dependent predicates are concerned.

\paragraph{Outline}
We first show the encoding of a verification problem using the example
of universal termination verification.  Then we discuss the challenges
associated with decomposing this problem and the interdependencies
with the solving process.

\vspace*{-0.9ex}
\section{Encoding}
\vspace*{-0.9ex}

We assume that programs are given in terms of call graphs, where
individual {\function}s $f$ are given in terms of symbolic
input/output transition systems.  Formally, the input/output
transition system of a {\function} $f$ is a triple of characteristic
predicates for relations $(\init_f,\trans_f,\fout_f)$, where
$\trans_f(\vx,\vxp)$ is the transition relation; the input relation
$\init_f(\vxin, \vx)$ defines the initial states of the transition
system and relates them to the inputs $\vxin$; the output relation
$\fout_f(\vx,\vxout)$ connects the transition system to the outputs
$\vxout$ of the {\function}.
Inputs are {\function} parameters, global variables, and memory
objects that are read by $f$.  Outputs are return values, 
global variables, and memory objects
written by $f$. Internal states $\vx$ are usually the values of
variables at the loop heads in $f$.
These relations are given as \emph{first-order logic formulae}
resulting from the logical encoding of the program semantics.

Let $F$ denote the set $\{f_1,\ldots,f_n\}$ of {\functions} in a given
program.  $H_f$ is the set of {\function} calls to {\functions} $h \in
F$ at calls sites $i$ in {\function} $f$. The vectors of input and
output arguments $\vxargin_{h_i}$ and $\vxargout_{h_i}$ are
intermediate variables in $\trans$.
We denote the termination argument $\rrank_f$, i.e.\ the conditions
that ensure the termination of {\function} $f$, such as a well-founded
transition invariant.

\paragraph{Example} 
By encoding Hoare-style verification rules (cf.~\cite{GLPR12}) into
second-order logic,%
\footnote{Mind that we use the notation $\exists_2$ to stress the fact
  that the quantifier binds a predicate.}
 we obtain the following formula. Its satisfiability
guarantees universal termination of the program.
\begin{equation}\label{equ:enc}
\small
\begin{array}{@{\hspace*{-1em}}lrl}
\multicolumn{3}{@{\hspace*{-1em}}l}{\exists_2 \ssummary_{f_1},\ldots,\ssummary_{f_n}: \bigwedge_{f\in F}} \\
& \multicolumn{2}{l}{\exists_2 \inv_f, \rrank_f: \forall \vxin_f, \vx_f, \vxp_f, \vxout_f:} \\
&& \init_f(\vxin_f,\vx_f) \Longrightarrow \inv_f(\vx_f)\\
& \wedge &\inv_f(\vx_f)\wedge\trans_f(\vx_f,\vxp_f) \wedge \bigwedge_{h_i \in H_f} \ssummary_h(\xargin_{h_i},\xargout_{h_i}) \Longrightarrow \inv_f(\vxp_f) \wedge \rrank_f(\vx_f,\vxp_f) \\
& \wedge & \init_f(\vxin_f,\vx_f)\wedge\inv_f(\vxp_f)\wedge \fout_f(\vxp_f,\vxout_f) \Longrightarrow \ssummary_f(\vxin_f,\vxout_f)
\end{array}
\end{equation}

\noindent In this formula,
recursive {\functions} produce cyclic dependencies of their 
 $\ssummary_f$ predicates.
If abstractions are used to lazily solve the formula, the invariant
$\inv_f$ and the termination argument $\rrank_f$ become
interdependent.
Similarly, invariants of nested loops are dependent on each
other. Rewriting nested loops into a single loop with invariant
$\inv_f$ only ''hides'' these dependencies as relational dependencies
between the loop variables.

\vspace*{-0.9ex}
\section{Decomposition}
\vspace*{-0.9ex}

We can decompose a formula that encodes a verification problem, such as
(\ref{equ:enc}) above, into a sequence of subproblems that are
solved by the synthesis engine.  The soundness of the analysis result
is ensured by (1) the soundness of the
analysis of individual subproblems, (2) the soundness of the
combination of the subproblem results, (3) and induction over the
decomposition hierarchy.
Decomposition causes the following issues:
(A) It may introduce additional interdependent predicates.
(B) The subproblems may be inference, and not verification problems;
hence their solving requires optimisation (like our synthesis engine)
instead of decision procedures.

\begin{figure}[t]
\begin{subfigure}{0.3\textwidth}
\begin{tikzpicture}[scale=0.95]
\node (invfo) at (0,0) {$\inv_f$};
\node (ccfo) at (-1,1) {$\callctx_f$};
\node (sumfo) at (1.5,1) {$\summary_f$};
\node (rrf) at (1.5,0) {$\rrank_f$};
\node (sumsho) at (1.5,-1) {$\summary_h$};
\node (ccho) at (-1,-1) {$\callctx_h$};
\draw[->] (invfo) -- (ccfo);
\draw[->] (sumfo) -- (invfo);
\draw[<->] (rrf) -- (invfo);
\draw[->] (invfo) -- (sumsho);
\draw[->] (ccho) -- (invfo);
\draw[dashed,->] (sumsho) -- (ccho);
\draw[blue, rotate=34] (-0.7,-0.15) ellipse (1.35 and 0.7);
\draw[darkgreen, rotate=34] (0.95,0.05) ellipse (1.35 and 0.7);
\draw[red] (1.5,0) ellipse (0.6 and 0.4);
\end{tikzpicture}
\caption{\label{fig:uniterm}
Interprocedural universal termination verification problem
}
\end{subfigure}
\hspace{1em}
\begin{subfigure}{0.65\textwidth}
\begin{tikzpicture}[scale=0.95]
\node (invfo) at (0,0) {$\inv^o_f$};
\node (ccfo) at (-1,1) {$\callctx^o_f$};
\node (sumfo) at (1.5,1) {$\summary^o_f$};
\node (rrf) at (2.75,0) {$\rrank_f$};
\node (sumsho) at (1.5,-1) {$\summary^o_h$};
\node (ccho) at (-1,-1) {$\callctx^o_h$};
\node (invfu) at (5.5,0) {$\inv^u_f$};
\node (ccfu) at (7,1) {$\callctx^u_f$};
\node (sumfu) at (4,1) {$\summary^u_f$};
\node (prefu) at (4,2) {$\precond^u_f$};
\node (sumshu) at (4,-1) {$\summary^u_h$};
\node (cchu) at (7,-1) {$\callctx^u_h$};
\draw[->] (invfo) -- (ccfo);
\draw[->] (sumfo) -- (invfo);
\draw[->] (rrf) -- (invfo);
\draw[->] (invfo) -- (sumsho);
\draw[->] (ccho) -- (invfo);
\draw[dashed,->] (sumsho) -- (ccho);
\draw[->] (invfu) -- (ccfu);
\draw[->] (sumfu) -- (invfu);
\draw[->] (prefu) -- (sumfu);
\draw[->] (invfu) -- (rrf);
\draw[->] (rrf) -- (sumfu);
\draw[->] (invfu) -- (sumshu);
\draw[->] (cchu) -- (invfu);
\draw[dashed,->] (sumshu) -- (cchu);
\draw[->] (invfu) to [bend left=15] (invfo);
\draw[blue, rotate=34] (-0.7,0.00) ellipse (1.35 and 0.7);
\draw[darkgreen, rotate=34] (0.95,0.05) ellipse (1.35 and 0.7);
\draw[darkgreen, rotate=-34] (3.6,3.1) ellipse (1.35 and 0.7);
\draw[blue, rotate=-33] (5.6,3.15) ellipse (1.5 and 0.75);
\draw[dashed,red] (4.1,0.4) ellipse (1.9 and 1.0);
\end{tikzpicture}
\caption{\label{fig:condterm}
Interprocedural sufficient preconditions for termination inference
}
\end{subfigure}
\caption{\label{fig:cyclic}
Dependent predicates in the encodings and decompositions
}
\end{figure}
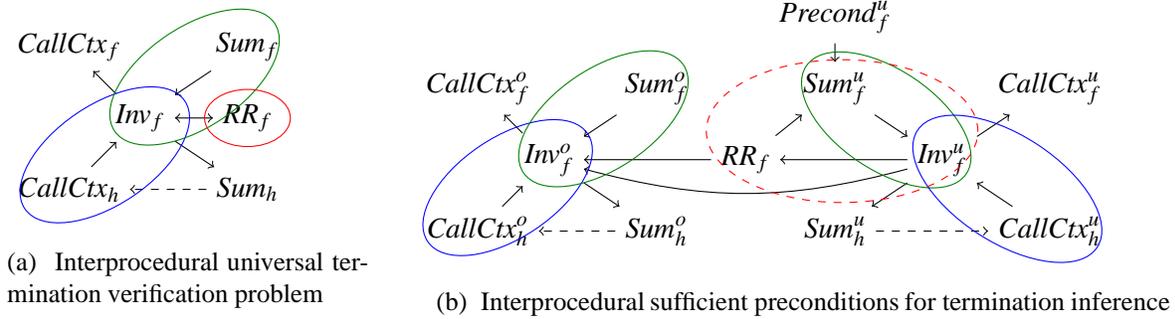

\paragraph{Example}
In \cite{CDK+15} we followed the classical approach of a procedural
decomposition. We emulate the traversal of the call graph top-down
analysing each {\function} separately and propagating the summaries
back up.  This decomposition splits the $\ssummary_{h}$ predicate for
a call to {\function} $h$ at call site $i$ into a \emph{calling
  context} predicate $\callctx_{h}$ that transfers information from
the caller to the callee,%
\footnote{ The calling context can be inferred by synthesising a
  predicate $\callctx_{h_i}$ s.t.  $\forall
  \vx_f,\vxp_f,\vxargin_{h_i},\vxargout_{h_i}: \inv_f(\vx_f) \wedge
  \trans_f(\vx_f,\vxp_f) \Longrightarrow
  \callctx_{h_i}(\vxargin_{h_i},\vxargout_{h_i})$.  } and a summary
predicate $\summary_{h}$ that transfers information from the callee to
the caller.  These two predicates are mutually dependent as
illustrated by the cycle in the dependency graph in
Figure~\ref{fig:uniterm}.  The dashed arrows are dependencies
resulting from unfolding the diagram along the call graph.  The blue,
green and red ellipses indicate the decomposition, i.e.\ predicates
that are solved for at once.
%
%
%
The algorithm 
in~\cite{CDK+15} uses a greatest fixed point
computation to resolve this dependency. However, this is very
imprecise for recursive {\functions}. 
%
%
%
%
%
Figure~\ref{fig:condterm} shows the predicate dependencies for the
inference of sufficient preconditions for termination.  Without going
into details (see~\cite{CDK+15}), we want to direct the attention to
the dependency (red dashed ellipse) between the
\emph{under-approximating} summary $\summary^u_f$ (of which the
sufficient precondition is a projection), the termination argument,
and the invariants -- which is a maximisation problem.


\vspace*{-0.9ex}
\section{Lessons Learned and Prospects}
\vspace*{-0.9ex}

%
We have to accommodate the following two conflicting goals: (1) Solving as large subformulae as possible to increase precision and
reduce the need for later refinement.  (2) Solving as small
subformulae as possible to be scalable. 
%
%
In Figure~\ref{fig:uniterm}, we solve for invariants (green) and
termination arguments (red) separately because our synthesis engine
currently does not support product domains that could infer both at
once, each with their optimised domains, thus eliminating cyclic
dependencies.
Some domains require least, others greatest fixed point computations,
our engine is currently unable to combine both in a single query.
Moreover, programs are rarely written with verification in mind; they
are often badly structured.  Therefore we need a property-,
precision-, and capacity-driven dynamic (de)composition to achieve
goals (1) and (2).  Re-partitioning the verification problem by
eliminating predicates and introducing new ones seems essential.
Decompositions introducing cyclic dependencies should only be used if
the solving capacity is exceeded.  On the other hand, precision can be
increased by expansion, i.e.\ unrolling of loops and inlining of
recursions, if the capacity allows it. Many of these issues are akin
to open problems in neighbouring areas of research, e.g.\ \cite{HW13}.

\vspace*{-0.9ex}
\bibliographystyle{eptcs}
\bibliography{biblio}
\end{document}